\documentclass{rspublic}
\usepackage{amsfonts, exscale}
\usepackage{amssymb, amsthm,amsmath}

\date{}

\newtheorem{Theorem}{Theorem}
\newtheorem{Lemma}{Lemma}
\newtheorem{Proposition}{Proposition}
\newtheorem{Corollary}{Corollary}

\newcommand{\reff}[1]{{\rm(\ref{#1})}}

\newcommand{\pa}{\partial}

\renewcommand{\leq}{\leqslant}
\renewcommand{\geq}{\geqslant}

\renewcommand{\theenumi}{\emph{\roman{enumi}}}

\makeatletter
\renewcommand{\@biblabel}[1]{#1}
\makeatother

\begin{document}
\title[On localization of pseudo-relativistic energy]
{On localization of pseudo-relativistic energy}
\author[ A.A. Balinsky, A.E.
Tyukov]{Alexander A. Balinsky, Alexey E. Tyukov}
\affiliation{Cardiff School of Mathematics, Cardiff University,
Senghennydd Road, Cardiff CF24 4AG, United Kingdom } \maketitle
\begin{abstract}{Localization of energy, Dirichlet Laplacian, Kato inequality}
We present a Kato-type inequality for bounded domains $\Omega \subset
{\mathbb R}^n$, $n \geq 2$.
\end{abstract}

\section{Introduction}\label{s1}

Hardy's inequality is an important tool in the study of the spectral
properties of partial differential equations. This inequality states
that for a function
 $f \in C_0^\infty ({\mathbb R}^n)$, $n \geq 3$
$$
\int_{{\mathbb R}^n} {|f(x)|^2 \over |x|^2} \, dx \leq \const
\int\limits_{{\mathbb R}^n} |\nabla f (x)|^2  \, dx.
$$
The corresponding "first order" analogue of the Hardy inequality
was established by Kato and  plays an important role in the study
of relativistic quantum mechanical systems. Specifically, Kato
inequality states that for $f \in C_0^\infty ({\mathbb R}^n)$, $n \geq
2$
\begin{equation}\label{50}
\int_{{\mathbb R}^n} {|f(x)|^2 \over |x|} \, dx \leq \const
\int\limits_{{\mathbb R}^n}  (\sqrt{-\Delta} f (x), f(x))  \, dx
\, ,
\end{equation}
where ${\Delta }=\sum\limits_{k=1}^n \pa^2_k$.
\par
The analogue of Hardy's inequality for  a bounded domain $\Omega \subset
{\mathbb R}^n$, $n \geq 2$ with  a Lipshitz boundary is
$$
\int\limits_{\Omega} {|u(x)|^2 \over (\rho_\Omega (x))^2} \, dx
\leq \const \int\limits_{ \Omega} |\nabla u(x)|^2 \, dx \, ,
$$
where $\rho_\Omega (x)=\min_{x_0 \in \pa \Omega}|x-x_0|$
(Edmunds\&Evans (2004) p. 212; see also Davies (1984, 1999) and
Lewis (1988) for references and details).
\par
The purpose of this article is to establish the Kato-type
inequality for a bounded domain $\Omega \subset {\mathbb R}^n$.
Since $\sqrt{-\Delta }$ is a non-local operator, there are three
possibilities to define the r.h.s. of~\reff{50} in the case of
$\Omega \subset {\mathbb R}^n$.  One possibility is to use
the r.h.s. of~\reff{50} but restrict ourselves  only to functions
with compact support inside $\Omega$. Another possibility is based
on the fact that (see Lieb\&Loss (1997))
$$
\int\limits_{{\mathbb R}^n}  (\sqrt{-\Delta} f (x), f(x))  \, dx
={\Gamma ({n-1 \over 2})\over 2 \pi^{n+1 \over 2}}
\int\limits_{{\mathbb R}^n \times {\mathbb R}^n} {|f(x)-f(y)|^2
\over |x-y|^{n+1}} \, dx dy\, .
$$
So we can define  the analogue of the r.h.s. of~\reff{50} for
$\Omega$ as
\begin{equation}\label{11}
{\Gamma ({n-1 \over 2})\over 2 \pi^{n+1 \over 2}}
\int\limits_{\Omega \times \Omega } {|f(x)-f(y)|^2 \over
|x-y|^{n+1}}  \, dx dy \, .
\end{equation}
The third possibility is to consider  square root of the internal
Dirichlet Laplacian operator in the domain ${\Omega}$.
\par
In this article we consider the first two definitions, since they
are more interesting for relativistic quantum mechanics
(localization of kinetic energy). The case of a Kato-type
inequality for the square root of the internal Dirichlet-Lapalcian
in fact follows for nice domains from Hardy's inequality since
$$
A^2 \geq B^2 \Longrightarrow A \geq B
$$
for  operators $A, B>0$ (see Birman\&Solomjak (1987), Theorem 2,
p. 232).
\par
Let us briefly describe the content of the paper.  In
section~\ref{s2} for functions $f$ such that $supp \, f \subset
\Omega_1$ for some $\overline \Omega_1 \subset \Omega$ we show
that
\begin{equation}\label{39}
\int\limits_{\Omega \times \Omega} {|f(x)-f(y)|^2 \over
|x-y|^{n+1}} \, dx dy \geq \const \int\limits_{\Omega_1} {|f
(x)|^2 \over  \rho_{\Omega_1} (x)} \, dx \, ,
\end{equation}
where $\rho_{\Omega_1} (x)$ is the distance from $x$ to $\pa
\Omega_1$, i.e. $ \rho_{\Omega_1} (x)= \min\limits_{z \in \pa
\Omega_1} |z-x|$. Later we obtain the inequality
\begin{equation}\label{58}
\int\limits_{\Omega \times \Omega } {|f(x)-f(y)|^2 \over
|x-y|^{n+1}} \, dx dy \geq \const \int\limits_{\Omega } {|f(x)|^2
\over \rho_{\Omega} ( x) (1+\left|\ln \rho_{\Omega} (x ) \right|^3
)} \, dx \, .
\end{equation}
Initially we prove~\reff{58} for radial functions
(Proposition~\ref{p1}) and then for all $f \in L^2(\Omega
,{\mathbb C})$ (Theorem~\ref{t2}). Though we give~\reff{58} for
some restricted class of bounded domains $\Omega$ we expect that
Theorem~\ref{t2} is true for more general domains. But we will not
discuss this in the current article.
\section{Kato-type inequality for functions with compact
support.}\label{s2}
\begin{Theorem}\label{t1}
Let $\Omega_1$ be a convex bounded domain such that
$\overline{\Omega}_1 \subset \Omega$ for some domain $\Omega
\subset {\mathbb R}^n$, $n \geq 2$. We suppose that $f \in
L^2(\Omega, {\mathbb C}^1)$ and $supp \, f \subset \Omega_1$. Then
for some constant $c_1 =c_1(\Omega, \Omega_1)>0$ the
inequality~\reff{39} holds.
\end{Theorem}
In view of the inequality $|f(x)-f(y)| \geq ||f(x)|-|f(y)||$,
without loss of generality we may assume that $f(x)$ is a
real-valued function. Next we apply the Lieb-Yau trick (see
Lieb\&Yau (1988)) to get  inequality which is a basic tool in the
proofs of Theorems~\ref{t1} and~\ref{t2}.
\begin{Lemma}\label{l4}
Let $K: B \times B \rightarrow {\mathbb R}$, $h: B \rightarrow
{\mathbb R}$, where $B \subset {\mathbb R}^m$, $m \in {\mathbb
N}$. We assume that $K \in L^\infty (B \times B)$,
\begin{equation}\label{23}
K(x, y)=K(y, x) \, , \qquad K(x,y) \geq 0 \,
\end{equation}
and
\begin{equation}\label{30}
0< M< h(z) <M^{-1}
\end{equation} for any $x,y,z \in B$ and some constant $M>0$. Then
\begin{equation}\label{22}
\int\limits_{B} \int\limits_{B}  {(f(x) -f(y))^2 } \, K(x, y) \,
dx dy \geq 2 \int\limits_{B} f^2(x) \, \bigg[\int\limits_{B}
K(x,y) \bigg(1-{h(x) \over h(y)} \bigg) \, dy \bigg] dx
\end{equation}
for any $f \in L^2 (B)$ with bounded support.
\end{Lemma}
\begin{proof}
On expanding  brackets in the l.h.s. of~\reff{22} we get
\begin{eqnarray}\label{25}
\lefteqn{ \int\limits_{B} \int\limits_{B}  {(f(x) -f(y))^2 } \,
K(x, y) \, dx dy} \qquad
\\[1mm] \nonumber
&\geq&   2 \int\limits_{B} f^2(x)\bigg[\int\limits_{B} K(x, y) \,
dy \bigg] \, dx -2 \int\limits_{B} \int\limits_{B} f(x) f(y) \,
K(x, y) \, dx dy \, .
\end{eqnarray}
Applying the Cauchy-Schwartz inequality  and using~\reff{23},
\reff{30} gives
\begin{eqnarray}\nonumber
\lefteqn{ \int\limits_{B} \int\limits_{B} f(x) f(y) \,
K(x, y) \, dx dy} \qquad \qquad\\[1mm] \nonumber &=& \int\limits_{B}
\int\limits_{B} \bigg(f(x) \sqrt{ K(x, y) h(x) \over h(y)} \,
\bigg) \, \bigg(f(y) \sqrt{ K(y, x) h(y) \over h(x)} \, \bigg)\,
dx dy
\\  \label{24} &\leq&
\int\limits_{B} \int\limits_{B} f^2(x) \, K(x, y) {h(x) \over
h(y)}\, dx dy \, .
\end{eqnarray}
The inequality~\reff{22} follows from~\reff{25} and \reff{24}.
\end{proof}
\begin{Corollary}\label{c2}
Let us suppose that $K(x, y)$ and $f(x)$ satisfy to conditions of
Lemma~\ref{l4}  and $supp \, f \subset \Omega_1$ for some
$\Omega_1 \subset B$. Then
\begin{equation}
\int\limits_{B} \int\limits_{B}  {(f(x) -f(y))^2 } \, K(x, y) \,
dx dy \geq 2 \int\limits_{\Omega_1} f^2(x) \, \bigg[\int\limits_{B
\setminus {\Omega_1}} K(x,y)
 \, dy \bigg] dx
\end{equation}
\end{Corollary}
\begin{proof}
An application of Lemma~\ref{l4} with
$$
h_\varepsilon (z)=
    \begin{cases}
         1 \qquad \textrm{if} \quad z \in \Omega_1 \\[1mm]
         1/\varepsilon \qquad \textrm{otherwise}
    \end{cases} \,
$$
gives
$$
\int\limits_{B} \int\limits_{B}  {(f(x) -f(y))^2 } \, K(x, y) \,
dx dy \geq 2 \int\limits_{\Omega_1} f^2(x) L_\varepsilon (x)\,  dx
\, ,
$$
where $$ L_\varepsilon (x)=\int\limits_{B} K(x,y)
\bigg(1-{h_\varepsilon (x) \over h_\varepsilon (y)} \bigg) \, dy =
\int\limits_{B \setminus {\Omega}} K(x,y) (1-\varepsilon) \, dy \,
.
$$
Passing to the limit $\varepsilon \rightarrow 0$ completes the
proof.
\end{proof}
\begin{proof}[Proof of Theorem~\ref{t1}.]
In view of Corollary~\ref{c2} it suffices to prove that
\begin{equation}\label{26}
\int\limits_{\Omega\setminus {\Omega_1}} {dy \over |x-y|^{n+1}}
\geq { c_2 \over \rho (x)} \,
\end{equation}
for any $x \in \Omega_1$ and some $c_2=c_2 (\Omega, \Omega_1)>0$.
The convexity of $\Omega_1$ implies that for any $z \in \pa
\Omega_1$ there exists an $(n-1)$-dimensional plane $\pi_{z}$ in
${\mathbb R}^n$ such that $ z \in \pi_z$ and $\pi_z \cap \Omega_1
= \emptyset$. For any $x \in \Omega_1$ we take $x_0=x_0 (x)$ such
that $\rho_{\Omega_1} (x)=|x-x_0|$. Let  $D_{x_0}$ be the half of
${\mathbb R}^n$ with boundary $\pi_{x_0}$ which does not contain
$\Omega_1$. Clearly
$$
\int\limits_{\Omega \setminus {\Omega_1}} { dy \over |x-y|^{n+1}}
\geq \int\limits_{\Omega \cap D_{x_0}} {dy \over |x-y|^{n+1}} \, .
$$
For any $z \in \pa \Omega_1$ we put
$$
\kappa_1 (z):= \sup\limits \{ s>0 : B_s (z) \subset \Omega\} \, ,
\qquad \kappa :=\inf\limits_{z \in \pa \Omega_1} k_1(z) \, ,
$$
where $B_s(z)$ is a ball with center at $z$ and radius $s$. From
$\overline{\Omega}_1 \subset \Omega$ we conclude that
$\kappa=\kappa(\Omega, \Omega_1)>0$. Consequently we have
\begin{equation}\label{28}
\int\limits_{\Omega \cap D_{x_0}} { dy \over |x-y|^{n+1}}
 \geq \int\limits_{B_\kappa (x_0) \cap D_{x_0}} { dy \over |x-y|^{n+1}}
\, .
\end{equation}
Let us choose  Cartesian coordinates $(y_1,\ldots,  y_n)$ in
${\mathbb R}^n$ with center at $x_0$ and axes such that
 $D_{x_0}=\{y: y_1 \geq 0\}$. Then $x=(-\rho_{\Omega_1} (x), 0)$ and
$B_\kappa (x_0)=\{y: y_1^2+\ldots +y_n^2\leq \kappa^2\}$. Making
the change of  variables $y_1=\rho_{\Omega_1} (x) z_1, \ldots ,
y_n=\rho_{\Omega_1} (x) z_n$ in~\reff{28} gives
$$
\int\limits_{B_\kappa (x_0)\cap D_{x_0}}  { dy \over |x-y|^{n+1}}
={ 1 \over \rho_{\Omega_1} (x)} \,\int\limits_{S_1}  {dz_1 \ldots
dz_n \over ((z_1+1)^2+z_2^2+ \ldots +z_n)^{n+1 \over 2}} \, ,
$$
where
$$
S_1=\{z=(z_1,\ldots,  z_n): z_1>0 \quad \textrm{and} \quad
z_1^2+\ldots+ z_n^2 \leq  \kappa^2 (\rho_{\Omega_1} (x))^{-2} \}
\, .
$$
Since $\Omega_1$ is bounded, it follows that for some constant
$c_3=c_3(\Omega_1)>0$
$$
\rho_{\Omega_1} (x) \leq c_3
$$
for all $x \in \Omega_1$. Therefore $\rho_{\Omega_1}^{-2} (x) \geq
c_3^{-2}$ and so
$$
S_2:=\{z=(z_1, \ldots, z_n): z_1>0 \quad \textrm{and} \quad
z_1^2+\ldots+z_n^2 \leq  \kappa^2 c_3^{-2}\} \subset  S_1 \, .
$$
Combining the above estimates we obtain~\reff{26} with
$$
c_2= \int\limits_{S_2} \, {dz_1 \ldots dz_n \over
((z_1+1)^2+z_2+\ldots +z_n^2)^{n+1 \over 2}} \,   .
$$
\end{proof}
\section{Lower estimate for the integral representation~\reff{11}. Case of
radial functions.}\label{s3}
\begin{Proposition}\label{p1}
We suppose that $f:{\mathbb R}^1 \rightarrow {\mathbb R}^1 $ and
$supp \, f \subset [0,1)$. Then for some absolute constant $c_4
>0$ we get
\begin{equation}\label{36}
\int\limits_{B_1 (0)\times B_1 (0)} {(f(|x|)-f(|y|))^2 \over
|x-y|^{n+1}} \, dx dy \geq c_4 \int\limits_{B_1 (0)} {(f(|x|))^2
\over (1-|x|) (1-(\ln (1-|x|))^3)} \, dx \, ,
\end{equation}
where $B_1 (0) \subset {\mathbb R}^n$, $n \geq 2$,  is a ball with
center at the origin and  radius $R=1$.
\end{Proposition}
Let us briefly outline the content of this section. The proof of
Proposition~\ref{p1} is preceded by proofs of some auxiliary
results. In Lemma~\ref{l2} we show that integral on the l.h.s.
of~\reff{36} is equivalent (up to multiplication by a constant) to
one-dimensional integral~\reff{53}. In order to estimate~\reff{53}
from below we apply the Lieb-Yau trick (Lemma~\ref{l4}) with test
function
$$
h(r) = 100- (1-r)^\omega
$$
for $\omega \in (0, 1/4)$ and then  integrate in $\omega$ both
sides of the obtained inequality. Lemmas~\ref{l1}, \ref{l3} are
needed to get lower estimate for the term $\int\limits_{B} K(x,y)
(1-{h(x) / h(y)} ) \, dy$ on the r.h.s. of~\reff{22}. At the end
of this section we piece together all the lemmas to establish
Proposition~\ref{p1}.
\begin{Lemma}\label{l2}
Under the conditions of Proposition~\ref{p1},  for some constant
\\$c_5=c_5 (n) >0$  we have
\begin{equation}\label{10}
c_5  \,  \textrm{\bf I} \leq \int\limits_{B_1 (0)\times B_1 (0)}
{(f(|x|)-f(|y|))^2 \over |x-y|^{n+1}} \, dx dy \leq  2^{3-2n}
\pi^{2n-3} \, c_5 \, \, \textrm{\bf I} \, ,
\end{equation}
where
\begin{equation}\label{53}
\textrm{\bf I} = \int\limits_0^1 \int\limits_0^1 {(f(r) -f(s))^2
\over (r-s)^2 }  \, \left( {rs \over r+s} \right)^{n-1}   \, dr ds
\, .
\end{equation}
\end{Lemma}
\begin{proof}
Let us change  the coordinates $x,y$ in the integral in~\reff{10}
to spherical coordinates $x=(r, \theta_1, \ldots,  \theta_{n-1})$,
$y=(s, \phi_1, \ldots, \phi_{n-1})$, where
$$
r, s \in [0,1], \qquad \theta_1, \ldots, \theta_{n-2} , \phi_1,
\ldots, \phi_{n-2}  \in [0, \pi] \, , \qquad \theta_{n-1} ,
 \phi_{n-1} \in [0, 2\pi) \, .
$$
We choose the direction of the axes in $y$-space such that the
direction of axis $\phi_1=\pi/2$ coincides with the vector $x$,
i.e the angle between $x$ and $y$ is equal to $\phi_1$ and so
$$
|x-y|^2=|x|^2+|y|^2-2 |x||y| \cos\phi_1 \, .
$$
Recall that the absolute value of the Jacobian of this change of
variables is equal to
$$
\left(r^{n-1} |\sin \theta_1|^{n-2} \ldots |\sin
\theta_{n-2}|^{1}\right) \times \left(s^{n-1} |\sin \phi_1|^{n-2}
\ldots |\sin \phi_{n-2}|^{1} \right) \, .
$$
It follows that
\begin{align*}
\int\limits_{B_1 (0)\times B_1 (0)}& {(f(|x|)-f(|y|))^2 \over
|x-y|^{n+1}} \, dx dy
\\
= &c_6 \, \int\limits_0^1 \int\limits_0^1 \int\limits_{0}^{\pi }
{(f(r) -f(s))^2 \over (r^2+s^2-2rs\cos \phi_1 )^{n+1 \over 2} } \,
(rs)^{n-1} |\sin \phi_1 |^{n-2} \, dr ds d\phi_1 \, ,
\end{align*}
where
$$
c_6= \int\limits_{0}^{\pi} |\sin \theta_1|^{n-2} \, d\theta_1
\left(\int\limits_{0}^{\pi} |\sin \theta_2|^{n-3} \,
d\theta_2\right)^2 \times \ldots \times
\left(\int\limits_{0}^{\pi} |\sin \theta_{n-2}| \, d\theta_{n-2}
\right)^2\, .
$$
Denote by $J(k)$ the Euler-type integral
\begin{equation}\label{56}
J(k):=\int\limits_{0}^{\pi} {|\sin \phi |^{n-2}d \phi \over
(k^2+\sin^2 (\phi/2) )^{n+1 \over 2}} = 2 \int\limits_{0}^{{\pi
\over 2}} {|\sin (2 \phi) |^{n-2}d \phi \over (k^2+\sin^2 \phi
)^{n+1 \over 2}} \, .
\end{equation}
Then using
$$
r^2+s^2-2rs\cos \phi_1=(r-s)^2+4rs \sin^2 (\phi_1 /2)
$$
we obtain
\begin{align}\nonumber
\int\limits_{B_1 (0)\times B_1 (0)}& {(f(|x|)-f(|y|))^2 \over
|x-y|^{n+1}} \, dx dy
\\
\label{55} &=  c_6  \, \int\limits_0^1 \int\limits_0^1 {(f(r)
-f(s))^2 (rs)^{n-1}  \over (4rs)^{n+1 \over 2} } \,
J\bigg({|r-s|\over 2\sqrt{rs}}\bigg) \, dr ds  \, .
\end{align}
From~\reff{56} and  the elementary inequality $2 |z| / \pi \leq
|\sin z| \leq |z|$ for $z \in [-\pi/2 , \pi/2]$ we find that
$$
2^{n-1} \left( {2 \over \pi}\right)^{n-2} \int\limits_{0}^{{\pi
\over 2}} {\phi^{n-2}d \phi \over (k^2+ \phi^2 )^{n+1 \over 2}}
\leq J(k)\leq 2^{n-1} \int\limits_{0}^{{\pi \over 2}} {\phi^{n-2}
d \phi \over (k^2+ (2 /\pi)^2 \phi^2 )^{n+1 \over 2}} \, .
$$
Since
$$
\int {\phi^{n-2} \over (1+\phi^2)^{{n+1 \over 2}}} \, d\phi
={\phi^{n-1} \over (n-1) (1+\phi^2)^{{n-1 \over 2}}} +\const\, ,
$$
it follows that
\begin{align*}
J(k) &\geq  {1 \over k^2} \, 2^{n-1} \left( {2 \over
\pi}\right)^{n-2} {(\pi /2k)^{n-1} \over (n-1) (1+(\pi
/2k)^2)^{{n-1 \over 2}}} \, ,
\\
J(k)& \leq {1 \over k^2} \,  2^{n-1} \left({\pi \over
2}\right)^{n-1} {(1/k)^{n-1} \over (n-1) (1+(1/k)^2)^{{n-1 \over
2}}} \,
\end{align*}
or
$$
{1 \over  k^2} \, { 2^{n-2} \pi \over (n-1) (k^2+(\pi/2)^2)^{n-1
\over 2}} \leq J(k)\leq {1 \over k^2} \,  {\pi^{n-1} \over
(n-1)(k^2+1)^{n-1 \over 2}} \, .
$$
An application of the elementary inequality
$$
k^2+\left({\pi \over 2}\right)^2 \leq \left({\pi \over 2}\right)^2
(k^2+1)
$$
implies that
$$
{1 \over  k^2} \,{ 2^{2n-3} \pi^{2-\pi} \over (n-1) (k^2+1)^{n-1
\over 2}} \leq J(k)\leq {1 \over k^2} \, {\pi^{n-1} \over
(n-1)(k^2+1)^{n-1 \over 2}}  \, .
$$
Hence
$$
J\bigg({|r-s|\over 2\sqrt{rs}}\bigg) \leq {4rs \over  (r-s)^2} \,
 \, {  \pi^{n-1}
(2\sqrt{rs})^{n-1} \over (n-1)(r+s)^{n-1 }}  =  { 2^{n+1}
\pi^{n-1} (rs)^{n+1 \over 2} \over (n-1)(r+s)^{n-1 }(r-s)^2}
$$
and
$$
J\bigg({|r-s|\over 2\sqrt{rs}}\bigg) \geq {4rs \over  (r-s)^2} \,
{ 2^{2n-3} \pi^{2-n}(2\sqrt{rs})^{n-1} \over (n-1) (r+s)^{n-1
\over 2}} = { 2^{3n-2} \pi^{2-n} (rs)^{n+1 \over 2} \over
(n-1)(r+s)^{n-1 }(r-s)^2} \, .
$$
Substituting these estimates into~\reff{55} we obtain
$$
 {2^{2n-3} \pi^{2-n}  \over
(n-1)} \,  c_6  \,  \textrm{\bf I} \leq \int\limits_{B_1 (0)\times
B_1 (0)} {(f(|x|)-f(|y|))^2 \over |x-y|^{n+1}} \, dx dy \leq {
\pi^{n-1} \over (n-1)}\,  c_6   \, \, \textrm{\bf I} \, .
$$
Taking $c_5 = 2^{2n-3} \pi^{2-n} c_6 / (n-1) \,   $ we arrive
at~\reff{10}.
\end{proof}
\begin{Lemma}\label{l1}
Let $\phi (\cdot)$ be  a positive increasing function and
\begin{equation}\label{31}
h(r)=\phi((1-r)^{-1}) \, .
\end{equation}
Then for any $r \in (0,1)$
\begin{eqnarray}\nonumber
\lefteqn{r^{-(n-1)} \lim\limits_{\varepsilon \rightarrow 0}
\int\limits_0^1 {(\min\{r, s\})^{n-1}  \over \varepsilon^2+(r-s)^2
} \, \bigg(1-{h (r) \over h (s)} \bigg) \, d s } \qquad \qquad
\\
\label{21} &\geq & \mu  \lim\limits_{\varepsilon \rightarrow 0}
\int\limits_{\mu^{-1}}^{+\infty}
 {1 \over
\varepsilon^2u^2+(u-1)^2 }  \, \bigg({ \phi (\mu u )- \phi ( \mu )
 \over \phi (\mu u )} \bigg) \, du  \, ,
\end{eqnarray}
where $\mu=(1-r)^{-1}$, $n \geq 2$.
\end{Lemma}
\begin{proof}
{\it Step 1.}
We have
\begin{align*}
\int\limits_0^1 (\min\{r, s\})^{n-1}  I  \, ds &=
\int\limits_0^{r} s^{n-1}  I  \, ds+r^{n-1}   \int\limits_{r}^1  I
\, ds \\ &= r^{n-1} \int\limits_{0}^1  I \, ds + \int\limits_0^{r}
(s^{n-1} -r^{n-1} ) I  \, ds \, ,
\end{align*}
where
$$
I={1  \over \varepsilon^2+(r-s)^2 }   \, \bigg(1-{h (r) \over h
(s)} \bigg) \, .
$$
Since $h(s)$ is increasing, then $I<0$ for $s<r$, and so
$$
\int\limits_0^{r} (s^{n-1} -r^{n-1} ) I  \, ds \geq 0 \, .
$$
Thus
\begin{align}\nonumber
\textrm{l.h.s. of~\reff{21}}&= \lim\limits_{\varepsilon
\rightarrow 0} r^{-(n-1)} \, \int\limits_0^1 (\min\{r, s\})^{n-1}
I \, ds
\\
\label{14} &\geq \lim\limits_{\varepsilon \rightarrow 0}
\int\limits_0^1 {1 \over \varepsilon^2+(r-s)^2 }   \, \bigg(1-{h
(r) \over h (s)} \bigg) \, ds \, .
\end{align}
\par {\it Step 2.}
Let us make the change of the variables
\begin{equation}\label{15}
u={ 1-r \over 1-s}
\end{equation}
in the integral on the r.h.s of~\reff{14}. Elementary calculations
give
\begin{align} \nonumber
s&= r+(1-r)\,\bigg(1- {1 \over u}\bigg) \, ,
\\[1mm]
\label{16} {1 \over 1-s} &={u \over 1-r} \, ,
\\[1mm] \label{17}
{1 \over \varepsilon^2 + (r-s)^2} &= {u^2 \over \varepsilon^2 u^2+
(1-r)^2 ( u-1)^2 } \, ,
\\[1mm]
\label{18} ds &=  { 1-r \over u^2 } \, du
\end{align}
and
\begin{equation}\label{19}
0 \leq s < 1 \qquad \Leftrightarrow  \qquad 1-r \, \leq u <
+\infty \, .
\end{equation}
Consequently, using~\reff{31} and~\reff{15}-\reff{19} we get
\begin{align}\nonumber
\textrm{r.h.s. of~\reff{14}} &= \lim\limits_{\varepsilon
\rightarrow 0} \int\limits_{1-r}^{\infty}{1-r \over \varepsilon^2
u^2+ (1-r)^2 ( u-1)^2 } \, \bigg(1-{ \phi \big({1\over 1-r} \big)
\over  \phi \big( {u \over 1-r} \big) } \bigg) \, du
\\  \label{20}
 &= {1 \over 1-r} \, \lim\limits_{\varepsilon \rightarrow
0} \int\limits_{1-r}^{\infty}{1 \over (1-r)^{-2}\varepsilon^2 u^2
+ ( u-1)^2 } \, \bigg(1-{\phi \big({1 \over 1-r} \big) \over  \phi
\big( {u \over 1-r} \big) } \bigg) \, du \, .
\end{align}
Substituting $\mu=(1-r)^{-1}$ into~\reff{20} and making the change
$\varepsilon:=\mu^2 \varepsilon $ we arrive at
\begin{equation}\label{27} \textrm{r.h.s. of~\reff{14}} \geq  \mu
\lim\limits_{\varepsilon \rightarrow 0} \int\limits_{\mu^{-1}
}^{+\infty}
 {1 \over
\varepsilon^2u^2+(u-1)^2 }  \, \bigg({ \phi (\mu u )- \phi ( \mu )
 \over \phi (\mu u )} \bigg) \, du \, .
\end{equation}
Combining~\reff{14} and~\reff{27} completes the proof.
\end{proof}
\begin{Lemma}\label{l3}
There exist absolute constants $c_7>0$ and $\kappa >0$ such that
for any $0 < \omega  < 1/4$ and $\mu >1$
\begin{equation}\label{13}
\lim\limits_{\varepsilon \rightarrow 0}
\int\limits_{\mu^{-1}}^{+\infty}
 {1 \over
\varepsilon^2u^2+(u-1)^2 }  \, \bigg({\phi (\mu u ) - \phi ( \mu )
 \over  \phi (\mu u )} \bigg) \, du
\geq {c_7 \, \omega^2  \over \mu^{\omega}} \, ,
\end{equation}
where
\begin{equation}\label{29}
\phi(z)=\kappa -z^{-\omega} \, .
\end{equation}
\end{Lemma}
\begin{proof}
Substituting~\reff{29} into the l.h.s of~\reff{13} and using
$$
{1 \over(\kappa-\mu^{-\omega}u^{-\omega})} ={1
\over(\kappa-\mu^{-\omega})} +
{\mu^{-\omega}u^{-\omega}-\mu^{-\omega} \over
(\kappa-\mu^{-\omega}) (\kappa-\mu^{-\omega}u^{-\omega})}
$$
we get
\begin{align}\nonumber
\textrm{l.h.s. of~\reff{13}} &= \lim\limits_{\varepsilon
\rightarrow 0} \, \int\limits_{\mu^{-1}}^\infty
{\mu^{-\omega}-\mu^{-\omega} u^{-\omega} \over (\varepsilon^2
u^2+(u-1)^2)(\kappa-\mu^{-\omega}u^{-\omega})} \, du
\\
\label{7} &= {\mu^{-\omega} \over \kappa-\mu^{-\omega}}\,A
 -  {\mu^{-2 \omega} \over
\kappa-\mu^{-\omega}}\, B
\end{align}
with
$$
A (\mu) =
 \lim\limits_{\varepsilon \rightarrow 0} \,
\int\limits_{\mu^{-1}}^{+\infty} {1-u^{-\omega} \over
\varepsilon^2 u^2 +(u-1)^2} \, du
$$
and
\begin{align}\nonumber
B (\mu) &= \lim\limits_{\varepsilon \rightarrow 0} \,
\int\limits_{\mu^{-1}}^{+\infty} {(u^{-\omega}-1)^2 \over
(\varepsilon^2u^2+(u-1)^2)(\kappa-\mu^{-\omega}u^{-\omega})} \, du
\\
\label{4} &= \int\limits_{\mu^{-1}}^{+\infty} {(u^{-\omega}-1)^2
\over (u-1)^2(\kappa-\mu^{-\omega}u^{-\omega})} \, du \, .
\end{align}
\par
Let us estimate $A(\mu)$ and $B(\mu)$. Since $\mu^{-1} \leq 1$ and
$1-u^{-\omega} <0$ for $u <1$, it follows that
\begin{equation}\label{1}
A(\mu) \geq   \lim\limits_{\varepsilon \rightarrow 0} \,
\int\limits_{0}^{+\infty} {1-u^{-\omega} \over \varepsilon^2
u^2+(u-1)^2} \, du  \, .
\end{equation}
For any $R \in (1, +\infty)$ we put
$$\gamma_R=\{u \in {\mathbb C}:
|u|=R,\,  \textrm{Im} \, u >0  \}\, ,  \qquad \gamma^1_R=[0,R]\, ,
\qquad \gamma^2_R=[-R, 0] \, .
$$
Let $\gamma_R$ be oriented anticlockwise and segments
$\gamma_R^1$, $\gamma_R^2$  oriented from left to right. Due to
the fact that for any $\varepsilon <1$
$$
\int\limits_{\gamma_R } {1-u^{-\omega}  \over \varepsilon^2 u^2 +
(u-1)^2} \, du \rightarrow 0 \quad \textrm{as} \quad R \rightarrow
+\infty
$$
an application of Cauchy's theorem gives
\begin{equation}\label{2}
\int\limits_{0}^{+\infty} \! \! \! {1-u^{-\omega}  \over
\varepsilon^2 u^2 +(u-1)^2} \, du = \! \! \lim\limits_{R
\rightarrow +\infty} \int\limits_{\gamma^2_R} {1-u^{-\omega} \over
\varepsilon^2 u^2 + (u-1)^2} \, du   = \! \!
\int\limits_{0}^{+\infty} {1-t^{-\omega}e^{-\ri \pi \omega} \over
\varepsilon^2 t^2 + (t+1)^2} \, dt  \, .
\end{equation}
Combining~\reff{1} and \reff{2} we have
$$
A(\mu) \geq \int\limits_0^{+\infty} {1-\cos(\pi \omega)
t^{-\omega}\over (t+1)^2} \, dt = (1-\cos(\pi \omega)) \,
\int\limits_0^{+\infty} { dt \over (t+1)^2} -\cos(\pi \omega) \psi
(\omega) \, ,
$$
where
$$
\psi (\omega) = \int\limits_0^{+\infty} {t^{-\omega}-1 \over
(t+1)^2} \, dt \, .
$$
Using the elementary inequality
$$
1-\cos\alpha \geq {\alpha^2 \over 4} \qquad \textrm{for} \qquad
\alpha \in \left[-{\pi \over 3}, {\pi \over 3}\right]
$$
we get
\begin{equation} \label{12}
A(\mu)  \geq {(\pi \omega)^2 \over 4} - \psi (\omega) \, .
\end{equation}
Making the change of the variables $t=z^{-1}$ we get
$$
\int\limits_{0}^{+\infty} {(\ln t)^{2m-1} \over (t+1)^2} \, dt = -
\int\limits_{0}^{+\infty} {(\ln z)^{2m-1} \over (z+1)^2} \, dz
$$
and so
\begin{equation}\label{59}
\psi^{(2m-1)} (0) = \int\limits_{0}^{+\infty} {(\ln t)^{2m-1}
\over (t+1)^2} \, dt =0 \qquad \textrm{for all} \quad m \in
{\mathbb N} \, .
\end{equation}
Moreover,
\begin{equation}\label{60}
\psi^{(2m)} (0) =\int\limits_{0}^{+\infty} {(\ln t)^{2m} \over
(t+1)^2} \, dt > 0 \qquad \textrm{for all} \quad m \in {\mathbb N}
\, .
\end{equation}
Applying~\reff{59}, \reff{60} to Taylor's expansion of
$\psi(\omega)$
$$
{\psi (\omega) \over \omega^2} = \sum\limits_{m=1}^\infty
{\psi^{(2m)} (\omega) \over (2m)!} \, \omega^{2(m-1)}
$$
we see that $\psi (\omega) \omega^{-2}$ increases for $\omega >0$ and so
$$
{\psi (\omega) \over \omega^2} \leq 16 \, \psi \left( {1 \over 4}
\right)
$$
for $0 < \omega < 1/4$. Therefore, by~\reff{12} we have
\begin{align}\label{5}
A (\mu) & \geq \left[ { \pi^2 \over 4} -16 \,  \psi \left( {1
\over 4} \right) \right] \omega^2 = c_8 \omega^2 \, ,
\end{align}
where
\begin{equation}\label{8}
c_8 = { \pi^2 \over 4} -16 \,\int\limits_0^{+\infty} {t^{-{1 \over
4}}-1 \over (t+1)^2} \, dt  \approx  0.695869349 \, .
\end{equation}
\par
We proceed with $B(\mu)$. According to~\reff{4}
$$
B(\mu) = \int\limits_{\mu^{-1}}^{+\infty} {(u^{-\omega}-1)^2 \over
(u-1)^2(\kappa-\mu^{-\omega}u^{-\omega})} \, du \leq {1 \over
\kappa -1} \,  \int\limits_{0}^{+\infty} {(u^{-\omega}-1)^2 \over
(u-1)^2} \, du \, .
$$
Since
$$
(u^{-\omega}-1)^2 \leq  \omega^2(\ln u )^2(u^{2 \omega}+u^{-2\omega})
\leq \omega^2(\ln u )^2(u^{1 \over 2}+u^{-{1 \over 2}})
$$
for all $ 0 < \omega <  1/4$ and all $u>0$, it follows that
\begin{equation}\label{6}
B(\mu) \leq {c_9 \over \kappa -1}\,  \omega ^2 \, ,
\end{equation}
where
\begin{equation}\label{9}
c_9=  \int\limits_{0}^{+\infty} {(\ln u )^2(u^{1 \over 2}+u^{-{1
\over 2}}) \over (u-1)^2} \, du \approx 39.47841761 \,.
\end{equation}
Combining~\reff{7} with \reff{5}, \reff{6} we have
\begin{align*}
\textrm{l.h.s. of~\reff{13}} =  {\mu^{-\omega} \over
\kappa-\mu^{-\omega}}\, \left( A
 -  \mu^{- \omega} \, B \right)
 &\geq
 {\mu^{-\omega}\omega^2  \over \kappa-\mu^{-\omega}}\, \left(  c_8
 -  \mu^{- \omega} \,{c_9 \over \kappa -1}\right)
\\[1mm]
 &\geq
 {\mu^{-\omega}\omega^2  \over \kappa+1}\,
 \left( c_8
 -  \,{c_9 \over \kappa -1} \right)
\end{align*}
for all $\mu>1$ and $0 < \omega < 1/4$. Taking $\kappa =100$ and
using~\reff{8}, \reff{9} we obtain~\reff{13} with
$$
c_7={1  \over \kappa+1}\,
 \left( c_8
 -  \,{c_9 \over \kappa -1} \right) \approx 0.002941558950 \, .
$$
\end{proof}
\begin{Lemma}\label{l6}
One has
$$
\int\limits_0^{1 \over 4} {\omega^2 \over \mu^\omega} \, d\omega
\geq {c_{10} \over 1+(\ln \mu)^3}
$$
for all $\mu \geq 1$ and some absolute constant $c_{10}>0$.
\end{Lemma}
\begin{proof}
After elementary calculations we get
\begin{equation}\label{40}
\psi(\mu):=\int\limits_0^{1 \over 4} {\omega^2 \over \mu^\omega}
\, d\omega= -{2+2\omega \ln \mu + \omega^2 (\ln \mu )^2 \over (\ln
\mu  )^3\mu^\omega} \bigg|_0^{1 \over 4}= {2 -\delta (\mu)\over
(\ln \mu )^3} \, ,
\end{equation}
where
$$
\delta (\mu):={2+2^{-1} \ln(\mu)+4^{-2} (\ln \mu )^2 \over \mu^{{1
\over 4} }} \, .
$$
Since
$$
\delta^\prime (\mu) = - { (\ln \mu)^2 \over 64 \mu^{5 \over 4}} <0
\, ,
$$
and so $2-\delta (\mu) \geq 2-\delta (e) $ for $\mu \geq e$, it
follows that
$$
\psi(\mu) \geq {2-\delta (e) \over (\ln \mu )^3} \geq {2-\delta
(e) \over 1+(\ln \mu )^3}
$$
for $\mu \geq e$. On the other hand, since $\psi^\prime
(\mu)=-\psi (\mu) (\ln \mu) <0$ we get
$$
\psi(\mu) \geq \psi(e) \geq {\psi (e) \over   1+(\ln \mu )^3}
$$
for $1 \leq \mu \leq e$. Note that, by~\reff{40}, \,  $\psi (e) =
2-\delta (e)$. Taking
$$
c_{10} =  2-\delta (e)= 2-{41 \over 16 \, e^{1 \over 4}} \approx
0.004322994
$$
we complete the proof.
\end{proof}
\begin{proof}[Proof of Proposition~\ref{p1}.]
Using the left inequality in~\reff{10} and the fact that
$$
{rs \over r+s} \geq {1 \over 2} \, \min\{r, s\}  \,
$$
we find
\begin{eqnarray}\nonumber
\lefteqn{ \int\limits_{B_1 (0)\times B_1 (0)} {(f(|x|)-f(|y|))^2
\over |x-y|^{n+1}} \, dx dy} \qquad \qquad
\\ \nonumber
&\geq& {c_5 \over 2^{n-1}} \int\limits_0^1 \int\limits_0^1 {(f(r)
-f(s))^2 \over (r-s)^2 } \, (\min\{r, s\})^{n-1} \, dr ds
\\
\label{32} &=& {c_5 \over 2^{n-1}} \lim\limits_{\varepsilon
\rightarrow 0}\int\limits_0^1 \int\limits_0^1 {(f(r) -f(s))^2
\over \varepsilon^2 +(r-s)^2 } \, (\min\{r, s\})^{n-1} \, dr ds \,
.
\end{eqnarray}
An application of Lemma~\ref{l4}  with $m=1$, $B=[0,1]$,
$$
K(r,s)= {(\min\{r, s\})^{n-1} \over \varepsilon^2+(r-s)^2}
$$
for any positive function $h(\cdot)$ gives
\begin{equation}\label{33}
\int\limits_0^1 \int\limits_0^1 {(f(r) -f(s))^2 \over
\varepsilon^2+ (r-s)^2 }  \, (\min\{r, s\})^{n-1}  \, dr ds \geq
\int\limits_0^1 (f(r))^2 L_\varepsilon (r) \, dr \, ,
\end{equation}
where
$$
L_\varepsilon (r)= 2 \, \int\limits_0^1 {(\min\{r, s\})^{n-1}
\over (\varepsilon^2+(r-s)^2)} \, \left(1-{h(r) \over h(s)}
\right) ds \, .
$$
Let $h(\cdot)$ and $\phi(\cdot)$ be defined by~\reff{31}
and~\reff{29} respectively. An application of Lemmas~\ref{l1}
and~\ref{l3} yields
\begin{align}\nonumber
\lim\limits_{\varepsilon \rightarrow 0} L_\varepsilon (r) &\geq 2
r^{n-1} \mu  \lim\limits_{\varepsilon \rightarrow 0}
\int\limits_{\mu^{-1}}^{+\infty}
 {1 \over
\varepsilon^2u^2+(u-1)^2 }  \, \bigg({ \phi (\mu u )- \phi ( \mu )
 \over \phi (\mu u )} \bigg) \, du
 \\
 \label{34}
 & \geq
 {2 c_7 r^{n-1} \mu \, \omega^2  \over \mu^{\omega}}  \, ,
\end{align}
where as before $\mu=(1-r)^{-1}$. Combining~\reff{32}-\reff{34} we
obtain
\begin{equation}\label{35}
\int\limits_{B_1 (0)\times B_1 (0)} {(f(|x|)-f(|y|))^2 \over
|x-y|^{n+1}} \, dx dy \geq {c_5 c_7 \over 2^{n-2}} \int\limits_0^1
{r^{n-1} \mu \, \omega^2  \over \mu^{\omega}} \, d r .
\end{equation}
Integrating both sides of~\reff{35} in $\omega$ and using
Lemma~\ref{l6} we have
\begin{eqnarray*}
\lefteqn{ \int\limits_0^{1 \over 4} 1 \, d \omega \,
\int\limits_{B_1 (0)\times B_1 (0)} {(f(|x|)-f(|y|))^2 \over
|x-y|^{n+1}} \, dx dy} \qquad \qquad \qquad \qquad \qquad \qquad
\\ &\geq& {c_5 c_7 \over 2^{n-2}} \int\limits_0^1 {r^{n-1} (f(r))^2
\over 1-r}  \,  \left( \int\limits_0^{1 \over
4} {\omega^2 \over \mu^{\omega}} \, d \omega \right) \, d r \\
&\geq& {c_5 c_7 c_{10}  \over 2^{n-2}} \int\limits_0^1 {r^{n-1}
(f(r))^2 \over (1-r) (1-(\ln (1-r))^3)} \, d r .
\end{eqnarray*}
We put
$$
c_4= {4 c_5 c_7 c_{10} \over 2^{n-2}c_{11}}  \, ,
$$
where
\begin{equation}\label{41}
c_{11}= \int\limits_{0}^{\pi} |\sin \theta_1|^{n-2} \, d\theta_1
\int\limits_{0}^{\pi} |\sin \theta_2|^{n-3} \, d\theta_2 \times
\ldots \times \int\limits_{0}^{\pi} |\sin \theta_{n-2}| \,
d\theta_{n-2}\, .
\end{equation}
Thus
\begin{equation}\label{49}
\int\limits_{B_1 (0)\times B_1 (0)} {(f(|x|)-f(|y|))^2 \over
|x-y|^{n+1}} \, dx dy\geq c_4 c_{11} \int\limits_0^1 {r^{n-1}
(f(r))^2 \over (1-r) (1-(\ln (1-r))^3)} \, d r \, .
\end{equation}
Let us substitute~\reff{41} into~\reff{49} and change  the
variables $(r, \theta_1, \ldots, \theta_{n-1})$ on the r.h.s.
of~\reff{49} to Cartesian coordinates. Then~\reff{36} follows.
\end{proof}
\section{Lower estimate for the integral representation~\reff{11}. General case.}
\label{s4} Here we generalize inequality~\reff{36} to the case of
non-radial functions. Furthermore we obtain the analogue
of~\reff{36} for  certain class of domains $\Omega$.
\begin{Lemma}\label{l5}
Let $f \in L^2 \left({\mathbb R}^n , {\mathbb C}^1 \right)$, $n
\geq 2$ such that $ $ $supp \, f \subset B_1 (0) $. Then
\begin{equation}\label{37}
\int\limits_{B_1 (0)\times B_1 (0)} {|f(x)-f(y)|^2 \over
|x-y|^{n+1}} \, dx dy \geq c_4\int\limits_{B_1 (0)} {|f(x)|^2
\over (1-|x|) (1-(\ln (1-|x|))^3)} \, dx \, ,
\end{equation}
where $c_4>0$ is the absolute constant from Proposition~\ref{p1}.
\end{Lemma}
\begin{proof}
In view of the inequality $|f(x)-f(y)| \geq ||f(x)|-|f(y)||$,
without loss of generality we may assume that $f(x)$ is
real-valued.
\par
For any $e \in S^n$ ($S^n$ is the unit sphere in ${\mathbb R}^n$)
we put
$$
T^e : {\mathbb R}^n \rightarrow {\mathbb R}^n \, , \qquad T^e
z=2(z, e)e-z \, ,
$$
i.e. $T^e$ is rotation in ${\mathbb R}^n$ around $e$ through angle
$\pi$. Obviously
\begin{equation}\label{45}
|T^e x - T^e y|= |x-y|
\end{equation}
for all $x, y \in {\mathbb R}^n$.
\par
Making the change of the variables $x:=T^e x$, $y:=T^e y$ and
using~\reff{45} and $|\textrm{det} \, T^e|=1$ we have
\begin{equation}\label{51}
\int\limits_{B_1 (0)\times B_1 (0)} {|f(x)-f(y)|^2 \over
|x-y|^{n+1} } \, dx dy = \int\limits_{B_1 (0)\times B_1 (0)}
{|f(T^e x)-f(T^e
 y)|^2 \over |x-y|^{n+1}} \, dx dy \, .
\end{equation}
According to the Cauchy-Schwartz inequality
$$
\int\limits_{S^n}  f(T^e x) f(T^e y)\, de \leq
\left(\int\limits_{S^n} |f(T^e x)|^2\,
 de\right)^{1\over2} \left(\int\limits_{S^n}  |f(T^e y)|^2\,
 de\right)^{1\over2}
$$
and so
\begin{align*}
\int\limits_{S^n}  |f(T^e x)&-f(T^e
 y)|^2 \, de
 \\
 &=
 \int\limits_{S^n}  |f(T^e x)|^2\, de+
 \int\limits_{S^n}  |f(T^e y)|^2\, de-
2\int\limits_{S^n}  f(T^e x) f(T^e y)\, de
\\
&\geq  (\psi(x)-\psi(y))^2 \, ,
\end{align*}
where
$$
\psi (x):= \left(\int\limits_{S^n}  |f(T^e x)|^2\,
 de\right)^{1\over2} \, .
$$
Using this and integrating~\reff{51} over all $e \in S^n$ gives
\begin{equation}\label{52}
|S^n| \, \int\limits_{B_1 (0)\times B_1 (0)} {|f(x)-f(y)|^2 \over
|x-y|^{n+1}} \, dx dy \geq \int\limits_{B_1 (0)\times B_1 (0)}
{|\psi(x)-\psi(y)|^2 \over |x-y|^{n+1}} \, dx dy \, .
\end{equation}
Note that $\psi (x)$ depends only on $|x|$. Hence we can apply
Proposition~\ref{p1}. It follows that
\begin{align*}
\int\limits_{B_1 (0)\times B_1 (0)} {|\psi(x)-\psi(y)|^2 \over
|x-y|^{n+1}} \, dx dy &\geq c_4\int\limits_{B_1 (0)} {|\psi (x)|^2
\over (1-|x|) (1-(\ln (1-|x|))^3)} \, dx \\
&= c_4\int\limits_{B_1 (0)} {\int\limits_{S^n}  |f(T^e x)|^2\,
 de \over (1-|x|) (1-(\ln (1-|x|))^3)} \, dx \, .
\end{align*}
Since
$$
\int\limits_{B_1 (0)}   {|f(T^e x)|^2
  \over (1-|x|) (1-(\ln (1-|x|))^3)} \, dx
  = \int\limits_{B_1 (0)}   {|f(x)|^2
  \over (1-|x|) (1-(\ln (1-|x|))^3)} \, dx
$$
for any $e \in S^n$, it follows that
\begin{equation}\label{54}
\int\limits_{B_1 (0)\times B_1 (0)}\! {|\psi(x)-\psi(y)|^2 \over
|x-y|^{n+1} } \, dx dy  \geq c_4 |S^n| \! \!  \! \int\limits_{B_1
(0)} {|f(x)|^2
  \over (1-|x|) (1-(\ln (1-|x|))^3)} \, dx \, .
\end{equation}
Combining~\reff{52}, \reff{54} we complete the proof.
\end{proof}
\begin{Theorem}\label{t2}
Let $\Omega$ be a domain in ${\mathbb R}^n$, $n\geq 2$. We assume
that there exist diffeomorphism
$$
\phi: B_1 (0) \rightarrow \Omega
$$
and  some constant $c_{12}=c_{12} (\Omega)>1$ such that for  all
$u \in B_1 (0)$  \, $ \nabla \phi (u) > 0$ and
\begin{equation}\label{46}
c_{12}^{-1} \leq \lambda_i (u) \leq c_{12} \qquad i=1, \ldots, n
\, ,
\end{equation}
where $\lambda_i (u) $ are eigenvalues of the matrix $ \nabla \phi
(u) $. Then for some  constant \\ $ c_{14}= c_{14} (\Omega)>0$ and
any $f \in L^2 (\Omega, {\mathbb C}^1)$ we have
$$ \int\limits_{\Omega
\times \Omega } {|f(x)-f(y)|^2 \over |x-y|^{n+1}} \, dx dy \geq
c_{14} \int\limits_{\Omega } {|f(x)|^2 \over \rho_{\Omega} ( x)
(1+\left|\ln \rho_{\Omega} (x ) \right|^3 )} \, dx \, ,
$$
where $\rho_{\Omega} (y):=\min\limits_{y_0 \in \pa \Omega} |y-y_0|
$.
\end{Theorem}
\begin{proof}
{\it Step 1.} \reff{46} and the fact that $\det\nabla \phi (u)  =
\lambda_{1} \ldots \lambda_n $ imply that
\begin{equation}\label{47}
c_{12}^{-n} \leq \left|\det\nabla \phi (u)  \right| \leq c_{12}^n
\, , \qquad  c_{12}^{-n} \leq
  \left|\det\nabla\phi^{-1} (x) \right|  \leq c_{12}^n \, ,
\end{equation}
for all $u \in B_1(0)$ and  $x \in \Omega$. Moreover, for all $u,v
\in B_1(0)$ an application the mean value theorem to $\psi_1
(\tau) =\phi (\tau u+(1-\tau )v)$,  $\tau  \in [0,1]$ gives
\begin{equation}\label{48}
|\phi(u)-\phi(v)|=|\psi_1(1)-\psi_1(0)| \leq \max\limits_{\tau \in
[0,1]} |\psi_1^\prime(\tau)| \leq  c_{12} |u -v| \, ,
\end{equation}
where we have used, by~\reff{46}
$$|\psi_1^\prime(\tau)|\leq \max\limits_{w \in B_1(0)} \|\nabla \phi (w)  \| \,
|u-v| \leq \max\limits_{w \in B_1(0)}  \max\limits_{i=1..n}
\lambda_i (w) \, |u -v| \leq c_{12} |u-v| \, .
$$
Similarly,  for all $u,v \in B_1(0)$ we obtain
\begin{equation}\label{57}
|u -v|=|\psi_2 (1)-\psi_2 (0)| \leq \max\limits_{\tau \in [0,1]}
|\psi_2^\prime(\tau)| \leq  c_{12} |\phi(u) -\phi(v)| \, ,
\end{equation}
where  $\psi_2 (\tau ) =\phi^{-1} (\tau \phi(u)+(1-\tau)\phi(v))$.
\par
As before we put  $\rho_{D } (x)=\min\limits_{z \in \pa D}|x-z|$
for any domain $D$. Given any $x \in \Omega$ we put $u =\phi^{-1}
(x)$ and take $u_0$,  $x_1$ are such that
$$
\rho_{B_1(0)} (u) = |u_0-u| \, , \qquad \rho_{\Omega } (x) =
|x_1-x| \, ,
$$
i.e. $u_0$,  $x_1$ deliver minima to corresponding functionals.
Applying~\reff{48} we get
$$
\rho_{B_1(0)} (u) = |u_0-u| \geq c_{12}^{-1} |\phi(u_0)-\phi(u)|
=c_{12}^{-1} |x_0-x| \, ,
$$
where $x_0=\phi (u_0) $. By choice of $x_1$ we have $| x_0-x| \geq
| x_1-x|$ and so
\begin{equation}\label{44}
\rho_{B_1(0)} (u) \geq c_{12}^{-1} | x_1-x| =c_{12}^{-1}
\rho_{\Omega } (x) \, .
\end{equation}
Along these lines using~\reff{57} we get
\begin{equation}\label{38}
\rho_{\Omega } (x)  \geq c_{12}^{-1}  \rho_{B_1(0)} (u) \, .
\end{equation}
\par
{\it Step 2.} Making the change of  variables $x= \phi(u)$, $ y=
\phi(v)$, using~\reff{47}-\reff{48} and letting $g=f\circ \phi$ we
get
\begin{eqnarray}\nonumber
\lefteqn{ \int\limits_{\Omega \times \Omega } {|f(x)-f(y)|^2
\over |x-y|^{n+1}} \, dx dy} \qquad \qquad \\
\nonumber &=&  \int\limits_{B_1 (0)\times B_1 (0)} {|g(u)-g(v)|^2
\over |\phi(u)-\phi(v)|^{n+1}} \, \left| \det \, \nabla \phi(u) \,
\det \, \nabla \phi(v) \right| \, d u d v
\\
\label{42} &\geq& c_{12}^{-(3n+1)} \int\limits_{B_1 (0)\times B_1
(0)} {|g(u)-g(v)|^2 \over |u -v|^{n+1}} \,
 d u d v \, .
\end{eqnarray}
An application of Lemma~\ref{l5} yields
$$
\int\limits_{B_1 (0)\times B_1 (0)} {|g(u)-g(v)|^2 \over |u
-v|^{n+1}} \,
 d u d v
 \geq
 c_4\int\limits_{B_1 (0)} {|g( u)|^2 \over \rho_{B_1(0)} (u)
(1-(\ln \rho_{B_1(0)} (u) )^3)} \, d u \, .
$$
Using~\reff{44}  we find (recall $\rho_{B_1(0)} (u) <1 $) that
\begin{equation}\label{3}
1-(\ln \rho_{B_1(0)} (u) )^3\leq 1 -(\ln  c_{12}^{-1}
\rho_{\Omega} (\phi (u) )
 )^3  \leq c_{13} (1+\left|\ln
\rho_{\Omega} (\phi (u) ) \right|^3 )
\end{equation}
for some $c_{13}=c_{13} (\Omega, c_{12})>0$. We apply~\reff{38},
\reff{3} and then again make the change of  variables  $x=
\phi(u)$ and use~\reff{47} to get
\begin{eqnarray}\nonumber
\lefteqn{ \int\limits_{B_1 (0)\times B_1 (0)} \! \! {|g(u)-g(v)|^2
\over |u -v|^{n+1}} \,
 d u d v}
 \qquad \qquad \qquad \qquad
 \\
 \nonumber
 &\geq& c_{12}^{-1}c_{13}^{-1} c_4\int\limits_{B_1 (0)} {|g(
u)|^2 \over \rho_{\Omega} (\phi (u)) (1+\left|\ln \rho_{\Omega}
(\phi (u) ) \right|^3 )} \, d u
\\
\label{43}
 &\geq& c_{12}^{-(n+1)}c_{13}^{-1} c_4\int\limits_{\Omega} {|f( x)|^2 \over \rho_{\Omega} (x)
(1+\left|\ln \rho_{\Omega} (x) \right|^3 )} \, d  x \, .
\end{eqnarray}
Combining~\reff{42}, \reff{43} and letting $c_{14}=
c_{12}^{-(4n+2)}c_{13}^{-1} c_4$ we complete the proof.
\end{proof}
\begin{acknowledgements}
We are grateful for financial support through EPSRC Grant RCMT090. We also
thank Professor W.D. Evans for valuable discussions.
\end{acknowledgements}



\end{document}